\documentclass[aps,prb,twocolumn,showpacs,showkeys,superscriptaddress,floatfix,10pt]{revtex4-2}
\usepackage{amsmath,amssymb,amsfonts,stmaryrd,wasysym,graphicx,multirow,color,textcomp,subfigure}
 \usepackage{url,booktabs}
 \usepackage[colorlinks=true,citecolor=blue,urlcolor=blue]{hyperref}
 \usepackage{color,soul} 
 \usepackage{bm,cancel}
 \usepackage[normalem]{ulem}
 \usepackage[dvipsnames]{xcolor}

\begin{document}

\title{Controllable magnetic domains in twisted trilayer magnets}
\author{Kyoung-Min Kim}
\email{kmkim@ibs.re.kr}
\affiliation{Center for Theoretical Physics of Complex Systems, Institute for Basic Science, Daejeon 34126, Korea}
\author{Moon Jip Park}
\email{moonjippark@ibs.re.kr}
\affiliation{Department of Physics, Hanyang Univercity, Seoul 04763, Republic of Korea}

\begin{abstract}
The use of moiré patterns to manipulate two-dimensional materials has facilitated new possibilities for controlling material properties. The moiré patterns in the two-dimensional magnets can cause peculiar spin texture, as shown by previous studies focused on twisted bilayer systems. In our study, we develop a theoretical model to investigate the magnetic structure of twisted trilayer magnets. Unlike the twisted bilayer, the twisted trilayer magnet has four different local stacking structures distinguished by the interlayer couplings between the three layers. Our results show that the complex interlayer coupling effects in the moiré superlattice can lead to the stabilization of rich magnetic domain structures; these structures can be significantly manipulated by adjusting the twist angle. Additionally, external magnetic fields can easily manipulate these domain structures, indicating potential applications in spintronics devices.
\end{abstract}

\keywords{moiré engineering, twisted trilayer, var der Waals magnets}
\pacs{75.70.Ak, 75.75.+a, 75.10.Hk}

\maketitle

\section{Introduction}

The manipulation of two-dimensional materials using moiré patterns has facilitated exciting possibilities for designing unique material properties \cite{Balents2020,Andrei2020,Huang2022,Chen2020,Kennes2021,He2021,doi:10.1063/5.0070163,doi:10.1063/5.0105405}. Recent experimental demonstrations of moiré magnets in twisted transition metal trihalides MX3 (M = Cr, Ru; X = Cl, Br, I) have shown the potential to manipulate nanoscale magnetic domain structures by adjusting the twist angle \cite{doi:10.1126/science.abj7478,Xu2022,Xie2022}. Additionally, theoretical studies indicated the emergence of magnetic skyrmions \cite{doi:10.1021/acs.nanolett.8b03315,PhysRevB.103.L140406,PhysRevResearch.3.013027,Akram2021,PhysRevB.104.014410,Ghader2022} and topological magnons \cite{2206.05264,PhysRevB.107.L020404} within the domain structure, which are promising for the development of spintronics \cite{HIROHATA2020166711,doi:10.1063/5.0072735,Coey2012-bg} and multiferroic materials with electrical tunability \cite{Fumega_2023}.

The moiré engineering of the van der Waals interface possesses great potential for extension to various heterostructures and multilayer systems. Examples of recent progress include nonmagnetic 2D materials, such as graphene \cite{Balents2020,Andrei2020}, transition metal dichalcogenides \cite{Huang2022}, and transition metal oxides \cite{Chen2020}, as well as magnetic 2D materials, such as transition metal trihalides \cite{doi:10.1126/science.abj7478,Xu2022,Xie2022} with multilayer moiré superlattices.

\begin{figure} [t!]
	\includegraphics[width=.48\textwidth]{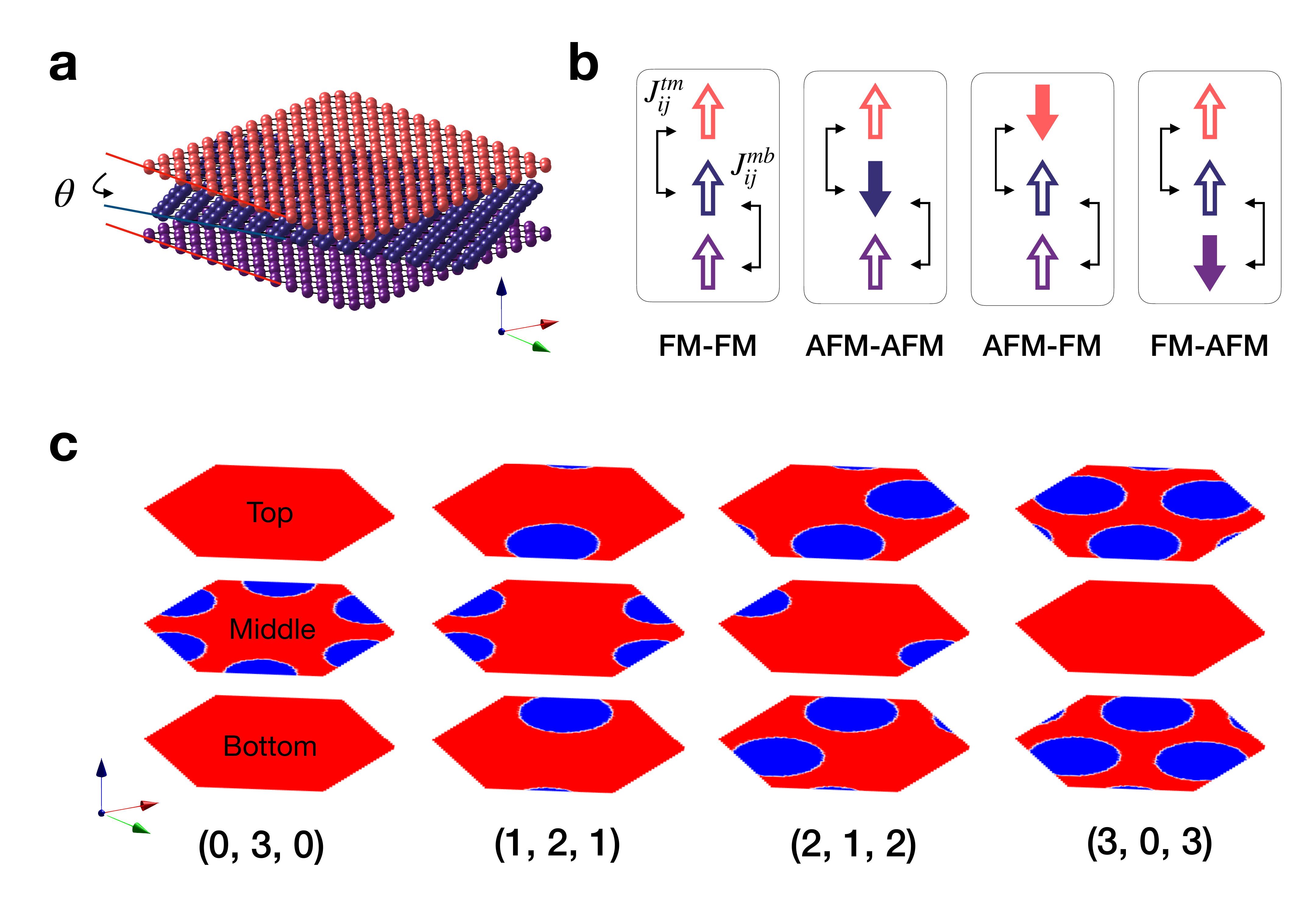}
    \caption{\textbf{Schematic of twist-engineering in trilayer magnets.} \textbf{a} Twisted trilayer magnet is formed by rotating the middle layer of an aligned trilayer magnet. \textbf{b} Twist generation of the four distinct local interlayer couplings $J_{ij}^\textrm{tm,mb}$, each with either ferromagnetic (FM) or antiferromagnetic (AFM) characteristics. \textbf{c} Interlayer coupling stabilization of the four different magnetic domain structures. Red (blue) indicates magnetization in the positive (negative) z-direction. The numbers in parentheses denote the number of magnetic domains per moiré unit cell in each layer.}
	\label{fig1}
\end{figure}

In this study, we present a theoretical model to investigate the effects of moiré patterns on twisted trilayer magnets (TTMs), as shown in Fig. \ref{fig1}\textbf{a}. In comparison to previous studies on twisted bilayer magnets (TBMs)\cite{Hejazi10721,PhysRevResearch.3.013027,Akram2021,PhysRevB.104.014410,2206.05264,Ghader2022,Fumega_2023,PhysRevLett.125.247201,https://doi.org/10.1002/adfm.202206923,PhysRevB.107.L020404}, the TTM has complex magnetic domain structures due to the four distinct coupling configurations of the local interlayer interaction between the layers, as shown in Fig. \ref{fig1}\textbf{b}. By examining the intriguing interlayer coupling effects in the moiré superlattice, we demonstrate that the magnetic domain structures can be altered by adjusting the twist angle and external magnetic field, as depicted in Fig. \ref{fig1}\textbf{c}. Our findings provide a controllable platform for magnetic textures and enable further exploration of the potential applications of moiré engineering in multilayer magnetic systems.

\section{Noncollinear magnetism induced by moiré patterns}

\begin{figure*}[t!]
\includegraphics[width=.96\textwidth]{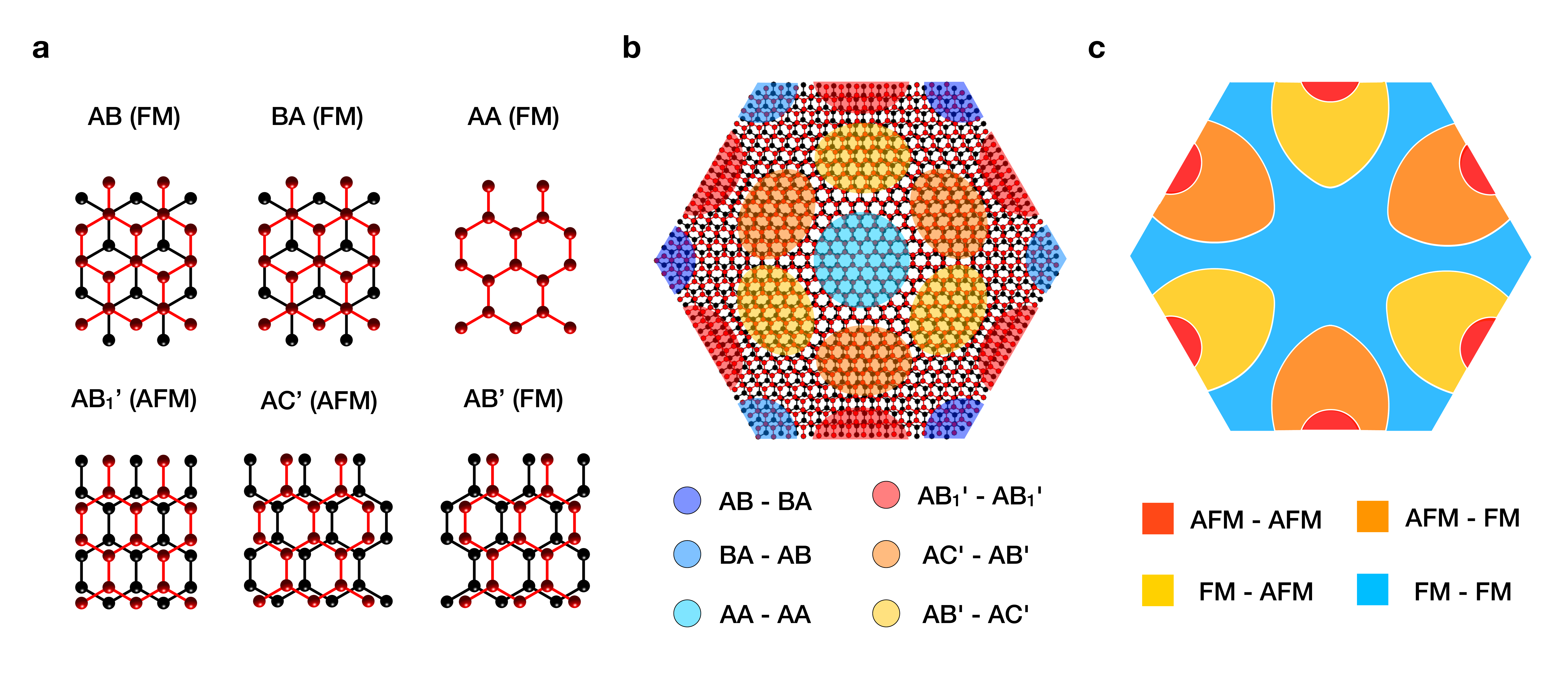}
\caption{\textbf{Local stacking patterns and interlayer exchange couplings.} \textbf{a} Two adjacent layers showing local stacking patterns that can be either ferromagnetic (FM) or antiferromagnetic (AFM), with corresponding interlayer exchange couplings. \textbf{b} Moiré unit cell with six distinct trilayer structures, each composed of the bilayer stacking patterns shown in \textbf{a}. The first (second) letters in each trilayer structure denote the bilayer stacking pattern between the top and middle (middle and bottom) layers. \textbf{c} Six local stacking regions are grouped into four patches with distinct interlayer coupling characteristics (such as AFM-AFM), where the first (second) letters indicate the coupling type of the interlayer exchange between the top-middle (middle-bottom) layers.}
\label{fig2}
\end{figure*}

The moiré superlattices of the TTMs are formed by rotating the middle layer of an aligned trilayer honeycomb lattice around one of its hexagonal centers via the commensurate angle $\theta_{m,n}=\arccos{\left(\frac{m^2+n^2+4mn}{2m^2+2n^2+2mn}\right)}$, where $m$ and $n$ are coprime integers \cite{PhysRevB.85.195458} [see Fig. \ref{fig1}\textbf{a}]. The Heisenberg spin model that describes the TTM is given by the following Hamiltonian,
\begin{align}
H &= \sum_{\textrm{l=t,m,b}}\left[-J \sum_{\langle i, j\rangle} \mathbf{m}_{i}^{\textrm{l}} \cdot \mathbf{m}_{j}^{\textrm{l}} - D_z\sum_{i} \big({m}_{z,i}^{\textrm{l}}\big)^2 \right] \nonumber \\
& + \sum_{i, j} \Big[ J_{ij}^{\textrm{tm}} \mathbf{m}_{i}^\textrm{t} \cdot \mathbf{m}_{j}^\textrm{m} + J_{ij}^{\textrm{mb}} \mathbf{m}_{i}^\textrm{m} \cdot \mathbf{m}_{j}^\textrm{b} \Big], \label{eq:spinH}
\end{align}
where $\mathbf{m}_{i}^{\textrm{l}}$ represents the magnetization vector for $i$-th spins in the top, middle, and bottom layers ($\textrm{l}=\textrm{t},\textrm{m},\textrm{b}$), respectively. $J>0$ is the intralayer FM exchange interactions between nearest-neighbor spins. $J_{ij}^{\textrm{l}_1,\textrm{l}_2}$ represents the interlayer exchange interactions between layer l$_1$ and l$_2$. $D_z>0$ indicates the single-ion anisotropy energy. 

In CrI$_3$, the stacking-dependent interactions alternatively exhibit FM and AFM interactions depending on the local stacking structures \cite{Huang2017,Sivadas2018}. To account for this behavior, we use the following coupling function, derived from ab initio calculations on laterally shifted bilayer CrI\textsubscript{3} \cite{2206.05264},
\begin{align}
J_{ij}^{\textrm{tm}, \textrm{mb}} & = J_0 \exp\big(-|r-d_0|/l_0\big) \nonumber \\
& + J_1^s \exp\big(-|r-r_*|/l_1^s\big) \sum_{a=1}^{3}\sin(\mathbf{q}_a^s \cdot \mathbf{r}_{ij}) \nonumber \\
& + J_1^c \exp\big(-|r-r_*|/l_1^c\big) \sum_{a=1}^{3}\cos(\mathbf{q}_a^c \cdot \mathbf{r}_{ij}), \label{eq:J_inter}
\end{align}
where $\mathbf{r}_{ij}=\mathbf{r}_i-\mathbf{r}_j$ represents the in-plane displacement between two interlayer spins, $d_0\approx6.7\textrm{\AA}$ is the out-of-plane distance, and $r=\sqrt{|\mathbf r_{ij}|^2+d_0^2}$. The wave vectors [$\mathbf{q}_{a}^{s,c}=q_{s,c}(\cos\varphi_a^{s,c},\sin\varphi_a^{s,c})$ for $a=1,2,3$, where $\varphi_a^{s}=(3\pi/2)(a-1)$ and $\varphi_a^{c}=(3\pi/2)(a-1)+\pi/6$], and length parameters ($l_0$, $l_1^{s,c}$) describe the leading oscillatory and decaying behavior of the interlayer coupling, respectively. The energy scales ($J_0, J_{1}^{s,c}$) determine the overall strength of the coupling. The specific values of the parameters are listed in Tab.\ref{tab1}.

The local stacking between neighboring layers exhibits continuous modulation between six distinct stacking configurations (AA, AB, BA, AB$'$, AC$'$, AB$_1'$). According to ab-initio calculations, AA, AB, BA, and AB$'$ stacking exhibit FM character, while AC$'$ and AB$_1'$ exhibit AFM character [See Fig.\ref{fig2}\textbf{a}] \cite{2206.05264}. In principle, the moiré superlattice of the TTM is composed of combinations of the six stacking structures. However, the stacking between the top-middle and the middle-bottom layers is not independent since the inversion symmetry ties the stacking structures between the top-middle and middle-bottom layers. As a result, the TTM possesses six local stacking structures: AB-BA, BA-AB, AA-AA, AB$_1'$-AB$_1'$, AC$'$-AB$'$, and AB$'$-AC$'$ [Fig. \ref{fig2}\textbf{b}], each of which has FM-FM, FM-FM, FM-FM, AFM-AFM, AFM-FM, or FM-AFM couplings for the top-middle and the middle-bottom layers, respectively. In turn, those local stacking regions can be grouped into four distinct local patches according to their interlayer coupling [See Fig.\ref{fig2}\textbf{c}].

\section{Magnetic domain structure via twist engineering}

Due to the coexistence of the AFM and FM interlayer couplings, the magnetic ground state forms a magnetic domain in one of the layers where the domain has an AFM spin configuration from its neighboring layer. In the TBM, the three $C_{3z}$ symmetric magnetic domains reside in one of the two identical layers, resulting in double degeneracy. In contrast, the TTMs possess multiple possibilities with distinct domain energies depending on the location of the domain. Specifically, the AFM-AFM patch can potentially stabilize the magnetic domains in the middle layer. Alternatively, the magnetic domain can span across a wider region, including the AFM-FM. FM-AFM patch, and the AFM-AFM patch, in either the top or bottom layer. In Fig. \ref{fig1}\textbf{c}, we classify the four possible distinct domain configurations, where the local interlayer coupling and the domain configurations are consistent with each other.

\begin{table}[!b]
\begin{tabular}{ccccccccc}
\toprule
$\textrm{J}_0(\textrm{meV})$ &$\textrm{J}^\textrm{s}_1('')$ & $\textrm{J}^\textrm{c}_1('')$ & $r_*(\textrm{\AA})$ &$l_0('')$ & $l^\textrm{s}_1('')$ & $l^\textrm{c}_1('')$ & $q_s(\textrm{\AA}^{-1})$ & $q_c('')$ \\ \midrule
$-0.1$ & $-0.5$ & $0.1$ & $7.3$ & $0.1$ & $0.3$ & $0.6$ & $0.7$ & $2.0$  \\\bottomrule
\end{tabular}
\caption{The parameters for the coupling function in Eq. \eqref{eq:J_inter}, which was derived from ab initio calculations on laterally shifted bilayer CrI\textsubscript{3} \cite{2206.05264}. }
\label{tab1}
\end{table}

\begin{figure*}[t!]
	\includegraphics[width=.96\textwidth]{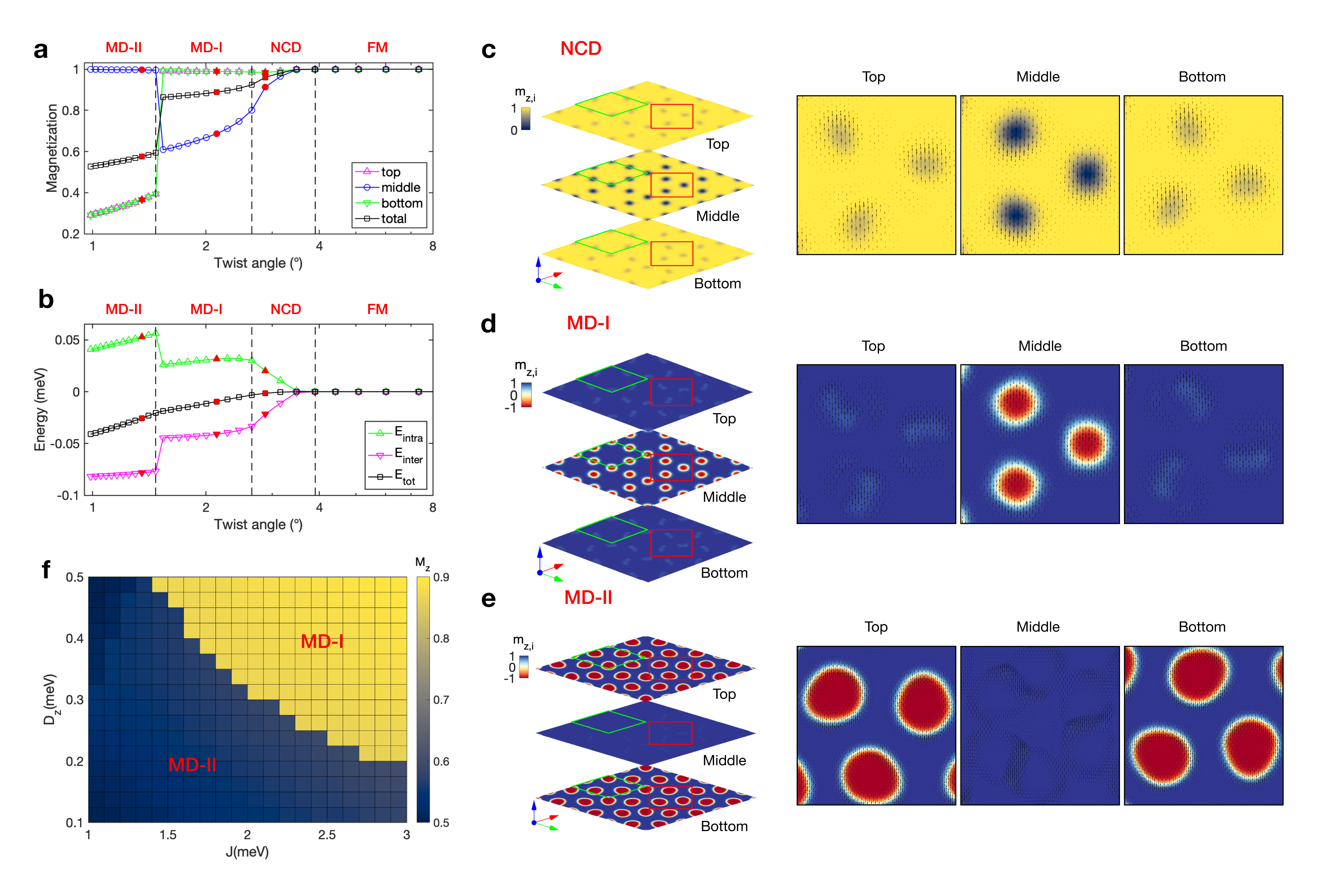}
   \caption{\textbf{Magnetic domain structures in twisted trilayer magnets.} \textbf{a} Phase diagram illustrating four magnetic phases: ferromagnetic (FM), noncollinear domain (NCD), type-I magnetic domain (MD-I), and type-II magnetic domain (MD-II) phases. The out-of-plane net magnetization ($M_z=\frac{1}{N}\sum_{i}m_{z,i}$) is shown as a function of twist angle ($\theta$) in the top (magenta), middle (blue), and bottom (green) layers, and all layers (black), respectively. \textbf{b} Magnetic energy per spin relative to the FM energy as a function of $\theta$. Each curve shows the intralayer gradient energy ($E_\textrm{intra}$), interlayer coupling energy ($E_\textrm{inter}$), and total energy ($E_\textrm{tot}=E_\textrm{intra}+E_\textrm{inter}$), respectively. \textbf{c} Magnetic configuration in the NCD phase, where the out-of-plane local magnetization ($m_{z,i}$) is depicted using a color scale. The $3\times3$ moiré superlattice unit cells are shown, with a green line indicating a single unit cell. The second to fourth panels depict the magnified area (red line) in each layer in the first panel, where the arrows denote the in-plane components of the local magnetization. \textbf{d} Magnetic configuration in the MD-I phase. \textbf{e} Magnetic configuration in the MD-II phase. \textbf{f} Phase diagram showing the transition from the MD-I to MD-II phases as a function of intralayer exchange ($J$) and single-ion anisotropy ($D_z$) with the out-of-plane net magnetization ($M_z$). For \textbf{a}-\textbf{e}, we use $J=2$ meV and $D_z=0.2$ meV. For \textbf{c}-\textbf{e}, the values of $\theta$ are denoted by red markers in \textbf{a} and \textbf{b}. For \textbf{f}, we use $\theta=2.45^\circ$ [$(m,n)=(13,14)$].}
	\label{fig3}
\end{figure*}

As the twist angle increases, the size of the moiré superlattice decreases. The magnetic domain structure is determined by the competition between the intralayer gradient energy ($E_\textrm{intra}$) and the interlayer coupling energy ($E_\textrm{inter}$). In the case of small superlattices, the intralayer gradient energy outweighs the interlayer coupling energy, resulting in a higher energy cost for magnetic domain formation. Therefore, it becomes energetically unfavorable for the system to form multiple magnetic domains. As a result, at large twist angles ($\theta\gtrsim 4^\circ$), the ground state exhibits a uniform FM order.

As the twist angle decreases, a continuous transition to the intermediate noncollinear domain (NCD) phase occurs before the formation of the magnetic domain. In the NCD phase, the spins in all layers are continuously tilted into the $x-y$ plane, resulting in a decrease in the net magnetization: $M_z=\frac{1}{3N}\sum_{\textrm{l=t,m,b}}\sum_{i=1}^N m_{z,i}^{\textrm{l}}$ [See Fig. \ref{fig3}\textbf{a} for the phase diagram]. This intermediate phase can be considered a precursor to the formation of the magnetic domain and is characterized by the continuous rotation of the magnetization vector, without a well-defined magnetic domain. The presence of the NCD phase is related to the interlayer exchange coupling between the top and middle layers, which competes with the intralayer exchange coupling and leads to the rotation of the spins.

As the twist angle decreases further, the spin tilt angle increases, resulting in a higher intralayer gradient energy $E_\textrm{intra}$ [Fig. \ref{fig3}\textbf{b}]. Despite this increase in $E_\textrm{intra}$, the interlayer coupling energy $E_\textrm{inter}$ results in a reduction of the total energy $E_\textrm{tot}=E_\textrm{intra}+E_\textrm{inter}$, stabilizing the magnetic domain (MD) phases where the spins in the layers flip inside the domain. Among the four possible categorizations of the MDs [Fig. \ref{fig1}\textbf{c}], two are energetically stabilized (MD-I and MD-II). In MD-I, a localized magnetic domain is found in the AFM-AFM patch of the middle layer; in the MD-II phase, the magnetic domains form in the top and bottom layers instead of the middle layer, doubling the number of domains from $3$ to $6$ per unit cell [Fig. \ref{fig3}\textbf{e}]. The number and size of domains are increased and include the AFM-FM and FM-AFM patches, as well as AFM-AFM patches, which in turn leads to a significant increase in the intralayer gradient energy compared to that in the MD-I phase [Fig. \ref{fig3}\textbf{b}]. However, the dominant interlayer coupling energy again results in a reduction in the total energy, leading to the stabilization of these extensive magnetic domains. Furthermore, this phase transition to the MD-II phase is a first-order transition and involves an abrupt reconstruction of the domain structure [Fig. \ref{fig3}\textbf{a}].

\begin{figure*}
    \centering
   \includegraphics[width=\textwidth]{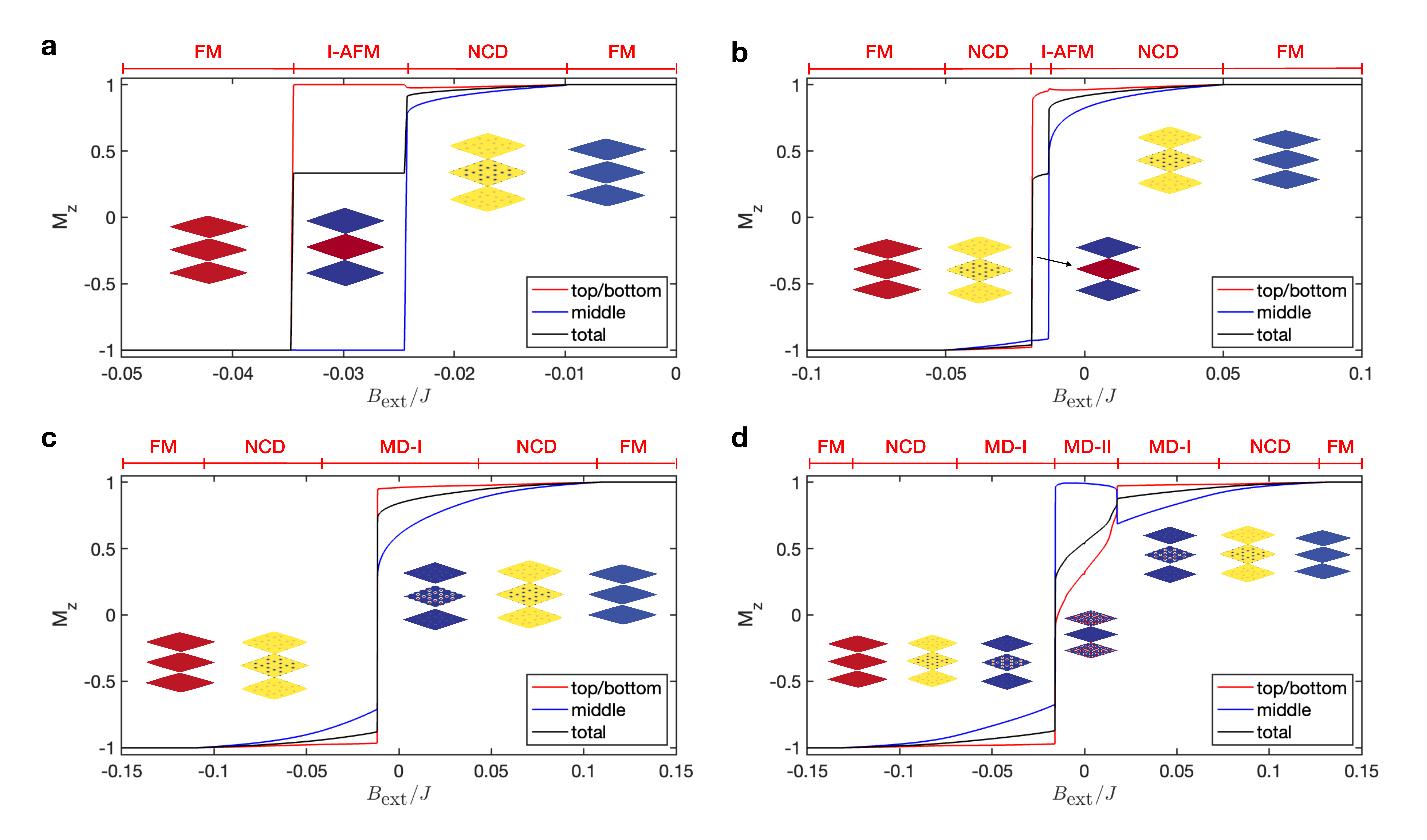}
    \caption{\textbf{Manipulating the magnetic domain structures with external magnetic fields.} \textbf{a} Effect of the external magnetic fields ($B_{\textrm{ext}}$) in the out-of-plane direction on a ferromagnetic (FM) ground state, transforming it into a noncollinear domain (NCD) or an interlayer antiferromagnetic (I-AFM) state. The net magnetization ($M_z$) is shown as a function of $B_{\textrm{ext}}$, where $B_{\textrm{ext}}M_z>0$ ($B_{\textrm{ext}}M_z<0$) indicates that the magnetic field is in the same (opposite) direction as the net magnetization. \textbf{b} Transformation of an NCD ground state into an FM or I-AFM state under the influence of a magnetic field. \textbf{c} Transformation of a type-I magnetic domain (MD-I) ground state into NCD or FM states. \textbf{d} Transformation of a type-II magnetic domain (MD-II) ground state into any of the previously mentioned magnetic states. The schematic images in each panel illustrate the corresponding magnetic structures. The values of $\theta$ used for \textbf{a}-\textbf{d} are $6.01^\circ$, $3.89^\circ$, $1.89^\circ$, and $1.30^\circ$, corresponding to (m,n)=(5,6),(9,10),(17,18),and (25,26), respectively. The parameters $J=2 \mathrm{meV}$ and $D_z=0.2 \mathrm{meV}$ are used for all plots.}
    \label{fig4}
\end{figure*}

\section{Manipulation of domain structure by applying external magnetic fields}

The magnetic domain structure of twisted trilayer magnets can be manipulated by applying external magnetic fields, without the need for fine-tuning the twist angle. Specifically, a uniform magnetic field applied in the out-of-plane direction ($B_{\textrm{ext}}$) can induce transformations between different magnetic states. To show this, we calculate stabilized magnetic structures modified from ground states in each magnetic phase by including the Zeeman coupling term,
\begin{equation}
H_Z = -B_{\textrm{ext}}\sum_{\textrm{l=t,m,b}}\sum_i m_{z,i}^\textrm{l},
\end{equation}
in addition to the magnetic Hamiltonian in Eq. \eqref{eq:spinH}.

We manipulate each magnetic phase through the applications of magnetic fields. First, as shown in Fig. \ref{fig4}\textbf{a}, an FM ground state can be transformed into an NCD state by applying a magnetic field in the opposite direction as the net magnetization ($M_z$). With increasing field strength, the NCD state can further evolve into a metastable state without domains, called an interlayer antiferromagnetic (I-AFM) state. In this state, the spins in the middle layer flip toward the field direction, while the other layers remain unchanged. As the field strength increases further, spins in the other layers also flip, resulting in the stabilization of an FM state in the direction of the magnetic field. 

Similarly, an NCD ground state can be transformed into an FM or I-AFM state by applying a magnetic field in the opposite direction [Fig. \ref{fig4}\textbf{b}]. Conversely, it can be continuously transformed into an FM state by applying a magnetic field in the same direction as the net magnetization. 

Moreover, an MD-I ground state can be continuously (or discontinuously with spin flipping) transformed into an NCD or FM state by applying a magnetic field in the same (opposite) direction as the net magnetization [Fig. \ref{fig4}\textbf{c}]. However, in this case, unlike the FM and NCD ground states, the I-AFM state is not stabilized. 

Finally, an MD-II ground state can be transformed into an MD-I state by applying a magnetic field in the same or opposite direction [Fig. \ref{fig4}\textbf{d}]. With increasing field strength, the MD-I state can further change into an NCD or FM state. As a result, all different magnetic structures of twisted trilayer magnets can be achieved by simply applying a magnetic field to an MD-II ground state, which offers wide flexibility in controlling the magnetic domain structure without requiring fine-tuning of the twist angle.

\section{Discussion}

Upon further investigation, we discovered that each magnetic phase could exhibit additional magnetic domain structures with varying numbers of small and large magnetic domains as excited states [see Fig. \ref{figS1}-\ref{figS3} in Appendix \ref{App3}], in addition to the ground magnetic configurations [Fig. \ref{fig3}\textbf{c}-\textbf{e}]. These excited states are notable for carrying skyrmion spin textures around the domain wall \cite{doi:10.1021/acs.nanolett.8b03315,PhysRevResearch.3.013027,Akram2021,PhysRevB.103.L140406,PhysRevB.104.014410,Ghader2022}, which can be stabilized due to the domain structure alone, without requiring other magnetic interactions that favor a winding spin texture, such as the Dzyaloshinskii-Moriya interaction \cite{Fert2013,Nagaosa2013,Fert2017}. These findings have significant implications since they demonstrate the possibility of manipulating magnetic domain structures to stabilize new states with unique spin textures, facilitating the development of novel spintronic devices with enhanced functionalities \cite{HIROHATA2020166711,doi:10.1063/5.0072735,Coey2012-bg}.

In addition, our findings suggest that the magnetic domains in twisted trilayer magnets can be easily manipulated via external magnetic fields, eliminating the need for precise control over the twist angle, which can be challenging in experimental techniques \cite{doi:10.1126/science.abj7478,Xu2022,Xie2022}. By applying a magnetic field, we can induce a transition between the different magnetic domain phases, enabling the precise control over the domain structure. Our study provides a promising avenue for designing magnetic devices based on twisted trilayer magnets, where the magnetic domains can be manipulated by external magnetic fields, enabling greater flexibility and control. \\

\acknowledgments{We acknowledge financial support from the Institute for Basic Science in the Republic of Korea through the project IBS-R024-D1. This work was also supported by the National Research Foundation of Korea(NRF) grant funded by the Korea government(MSIT) (RS-2023-00252085,RS-2023-00218998). M.J.P. would like to acknowledge the hospitality of the Center for Theoretical Physics of Complex Systems during the visit. Lastly, we would like to express our gratitude to Hee Chul Park for the helpful discussions.}

\bibliography{reference}

\appendix

\section{Iterative optimization method} \label{App1}

To obtain a stable magnetic configuration, we iteratively update a trial configuration by subtracting each magnetization vector $\mathbf{m}_i$ with the gradient vector of the magnetic Hamiltonian $H$. The update rule is given by,
\begin{equation}
\mathbf{m}_i^{(s+1)} = \mathbf{m}_i^{(s)} - c \mathbf{h}_i^{(s)},
\end{equation}
where $\mathbf{h}_i \equiv \frac{\partial H[{\mathbf{m}_i^{(s)}}]}{\partial\mathbf{m}_i^{(s)}}$ depends on the magnetic configuration ${\mathbf{m}_i^{(s)}}$ in each iteration step $(s)$, and $c$ controls the update rate. To preserve the norm of the magnetization vectors, we normalize each vector after each update, i.e., $\mathbf{m}_i^{(s+1)}\rightarrow \mathbf{m}_i^{(s+1)}/ \left|\mathbf{m}_i^{(s+1)}\right|$. We use periodic boundary conditions during the iterative updates to ensure consistency along the moiré unit cell boundary. We repeated this process more than 50 times with different trial initial configurations to determine the magnetic ground state, which is obtained by selecting the configuration with the lowest energy.

\section{Estimation of intralayer gradient and interlayer coupling energy} \label{App2}

We calculated the intralayer gradient energy $E_\textrm{intra}$ and interlayer coupling energy $E_\textrm{inter}$ presented in Fig. 3\textbf{b} of the main text by using the following equations:
\begin{align}
E_\textrm{intra}&=\frac{1}{N}(E_{J}+E_{D_z}-E_{J}^\textrm{FM}-E_{D_z}^\textrm{FM}),\\
E_\textrm{inter}&=\frac{1}{N}(E_{J_\perp}-E_{J_\perp}^\textrm{FM}),
\end{align}
where $E_{J}$, $E_{D_z}$ and $E_{J_\perp}$ denote the magnetic energy contribution from each magnetic interaction, defined as
\begin{align}
E_{J}&=-J\sum_{\textrm{l=t,m,b}}\sum_{\langle i, j\rangle} \mathbf{m}_{i}^{\textrm{l}} \cdot \mathbf{m}_{j}^{\textrm{l}},\\
E_{D_z}&=- D_z\sum_{\textrm{l=t,m,b}}\sum_{i} \big({m}_{z,i}^{\textrm{l}}\big)^2,\\
E_{J_\perp}&=\sum_{i,j} \Big[ J_{ij}^{\textrm{tm}} \mathbf{m}_{i}^\textrm{t} \cdot \mathbf{m}_{j}^\textrm{m} + J_{ij}^{\textrm{mb}} \mathbf{m}_{i}^\textrm{m} \cdot \mathbf{m}_{j}^\textrm{b} \Big],
\end{align}
and $E_{J}^\textrm{FM}$, $E_{D_z}^\textrm{FM}$, and $E_{J_\perp}^\textrm{FM}$ are the corresponding magnetic energies for a ferromagnetic state $\{\mathbf{m}_{i}^\textrm{l}\}=(0,0,1)$, respectively. $N$ denotes the total number of spins in the moiré unit cell, which is given by $N=\sum_{\textrm{l=t,m,b}}\sum_{i}$.

\section{Stabilized magnetic states with skyrmion spin textures} \label{App3}

\begin{figure*}
	\includegraphics[width=\textwidth]{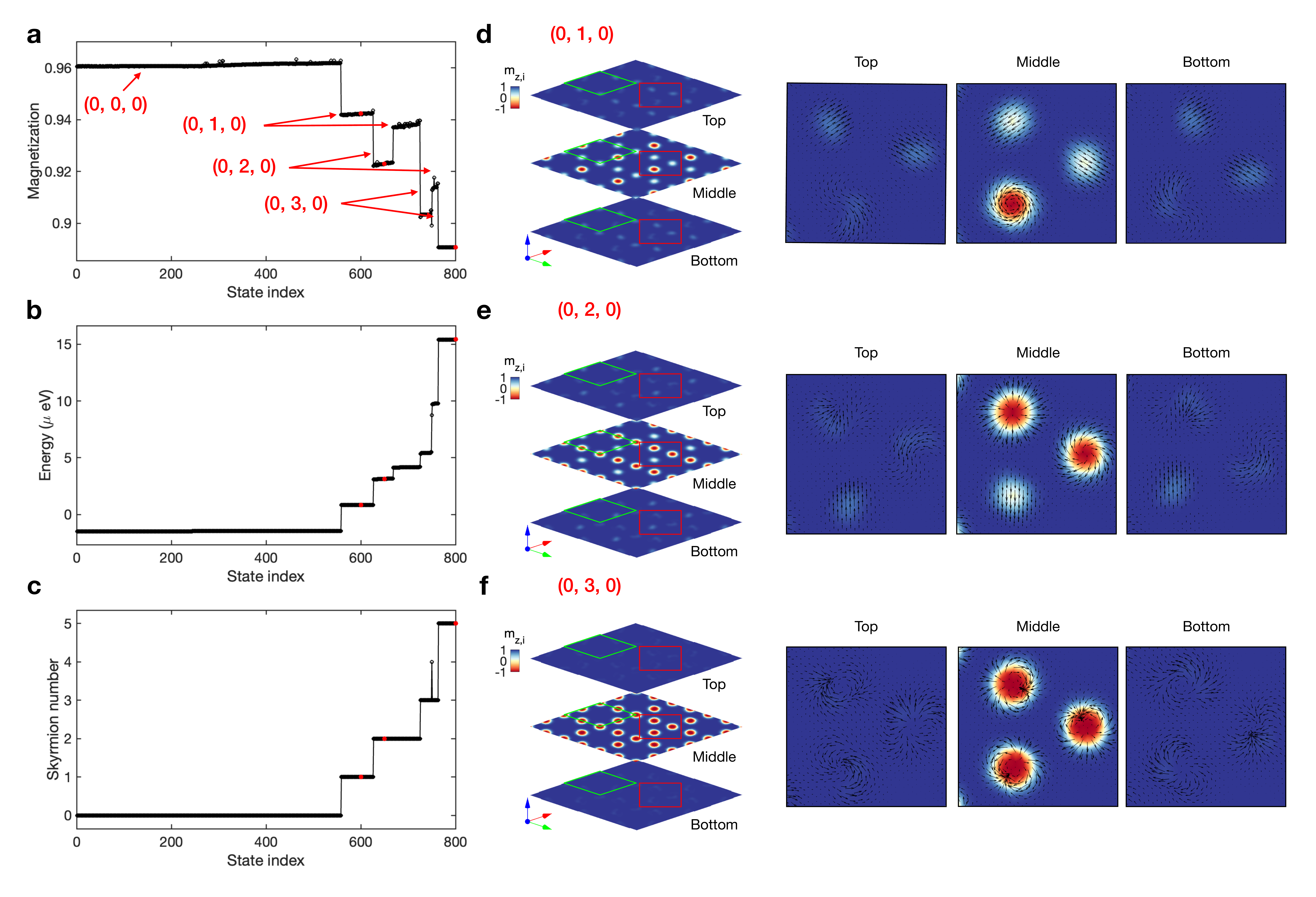}
    \caption{\textbf{Stabilized magnetic states and skyrmion spin textures in the noncollinear domain phase.} \textbf{a}-\textbf{c} Out-of-plane magnetization (\textbf{a}), magnetic energy per spin (\textbf{b}), and skyrmion number (\textbf{c}), where the state index denotes stabilized magnetic states from different random initial configurations sorted by energy. In \textbf{a}, the numbers in parentheses indicate the number of magnetic domains in each layer (top, middle, and bottom, respectively). \textbf{c} Magnetic configuration in the excited state with (0,1,0) domains, where the local magnetization in the out-of-plane direction ($m_{z,i}$) is depicted by a color scale. The $3\times3$ moiré superlattice unit cells are shown, where the green line denotes a single unit cell. The second to fourth panels show the magnified area (red line) in each layer in the first panel, where the arrows denote the in-plane components of the magnetization. \textbf{e} Magnetic configuration in the excited state with (0,2,0) domains. \textbf{f} Magnetic configuration in the excited state with (0,3,0) domains. For \textbf{a}-\textbf{f}, we use $J=2$ meV, $D_z=0.2$ meV, and $\theta=2.88^\circ$ [$(m,n)=(11,12)$].}
	\label{figS1}
\end{figure*}

Figure \ref{figS1} illustrates stabilized magnetic states obtained from various random initial configurations in the noncollinear domain (NCD) phase. In addition to the ground state with no domains [(0,0,0) state], magnetic domain states carrying different numbers of domains [(0,1,0), (0,2,0), (0,3,0)] are also identified, each having a reduced net magnetization [Fig. \ref{figS1}\textbf{a}]. The creation of domains increases the magnetic energy, indicating that they are metastable states [Fig. \ref{figS1}\textbf{b}]. These magnetic domains also contain skyrmion spin textures around their domain wall boundaries [Fig. \ref{figS1}\textbf{c}-\textbf{f}], where an increase in the skyrmion number results in a slight rise in energy.

\begin{figure*}
	\includegraphics[width=\textwidth]{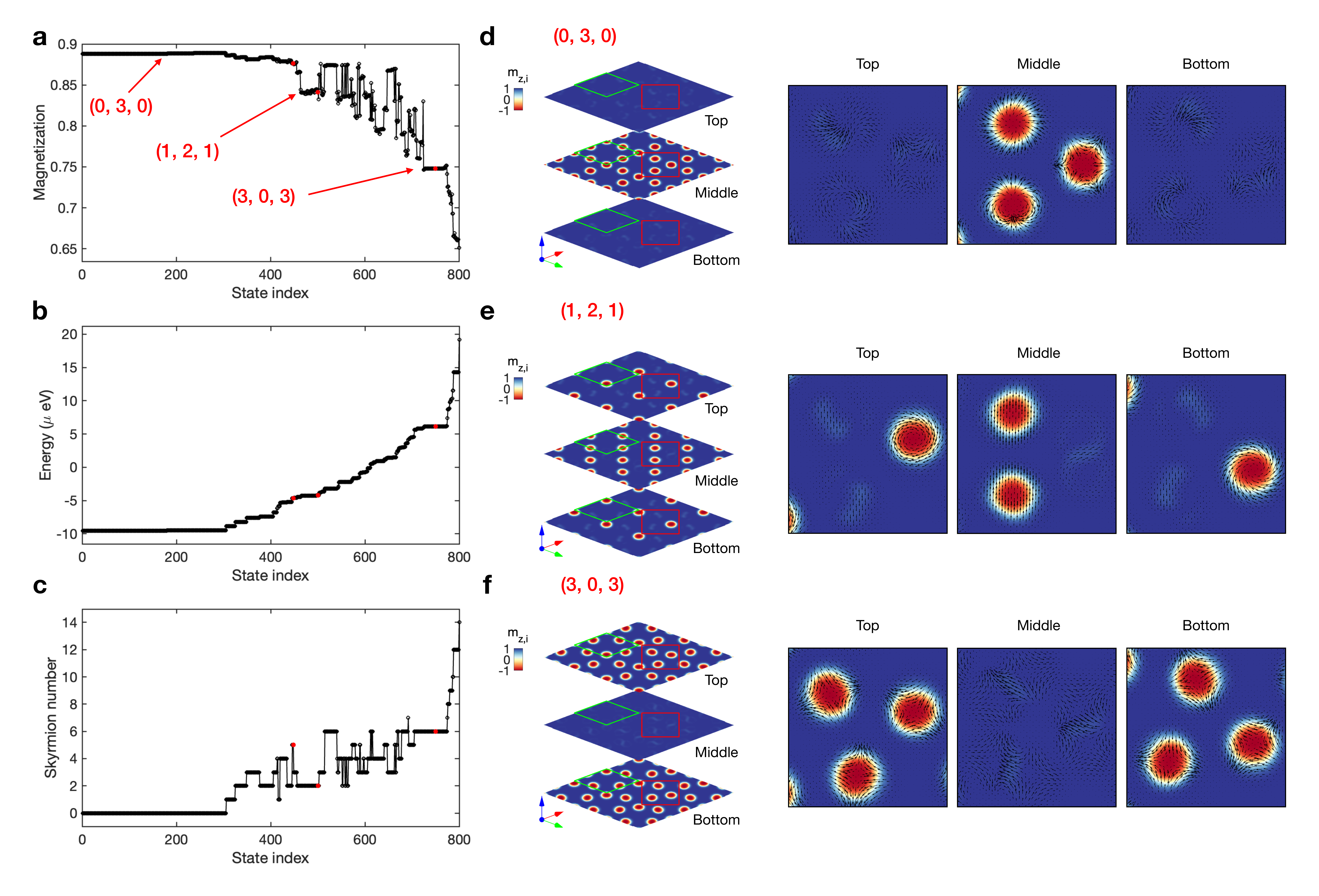}
     \caption{\textbf{Stabilized magnetic states and skyrmion spin textures in the type-I magnetic domain phase.} \textbf{a}-\textbf{c} Out-of-plane magnetization (\textbf{a}), magnetic energy per spin (\textbf{b}), and skyrmion number (\textbf{c}), where the state index denotes stabilized magnetic states from different random initial configurations sorted by energy. In \textbf{a}, the numbers in parentheses indicate the number of magnetic domains in each layer (top, middle, and bottom, respectively). \textbf{c} Magnetic configuration in the excited state with (0,3,0) domains, where the local magnetization in the out-of-plane direction ($m_{z,i}$) is depicted by a color scale. The $3\times3$ moiré superlattice unit cells are shown, where the green line denotes a single unit cell. The second to fourth panels show the magnified area (red line) in each layer in the first panel, where the arrows denote the in-plane components of the magnetization. \textbf{e} Magnetic configuration in the excited state with (1,2,1) domains. \textbf{f} Magnetic configuration in the excited state with (3,0,3) domains. For \textbf{a}-\textbf{f}, we use $J=2$ meV, $D_z=0.2$ meV, and $\theta=2.13^\circ$ [$(m,n)=(15,16)$].}
	\label{figS2}
\end{figure*}

Similarly, Figure \ref{figS2} illustrates stabilized magnetic states obtained from various random initial configurations in the type-I magnetic domain (MD-I) phase. In addition to the ground state with (0,3,0) domains, magnetic domain states carrying different numbers of domains [(0,1,0), (0,2,0), (3,0,3)] are identified, each having a reduced net magnetization [Fig. \ref{figS2}\textbf{a}]. The creation of domains leads to an increase in magnetic energy, indicating that they are metastable states [Fig. \ref{figS2}\textbf{b}]. Additionally, these magnetic domains contain skyrmion spin textures around their domain wall boundaries [Fig. \ref{figS2}\textbf{c}-\textbf{f}], where an increase in the skyrmion number results in a slight rise in energy.

\begin{figure*}
	\includegraphics[width=\textwidth]{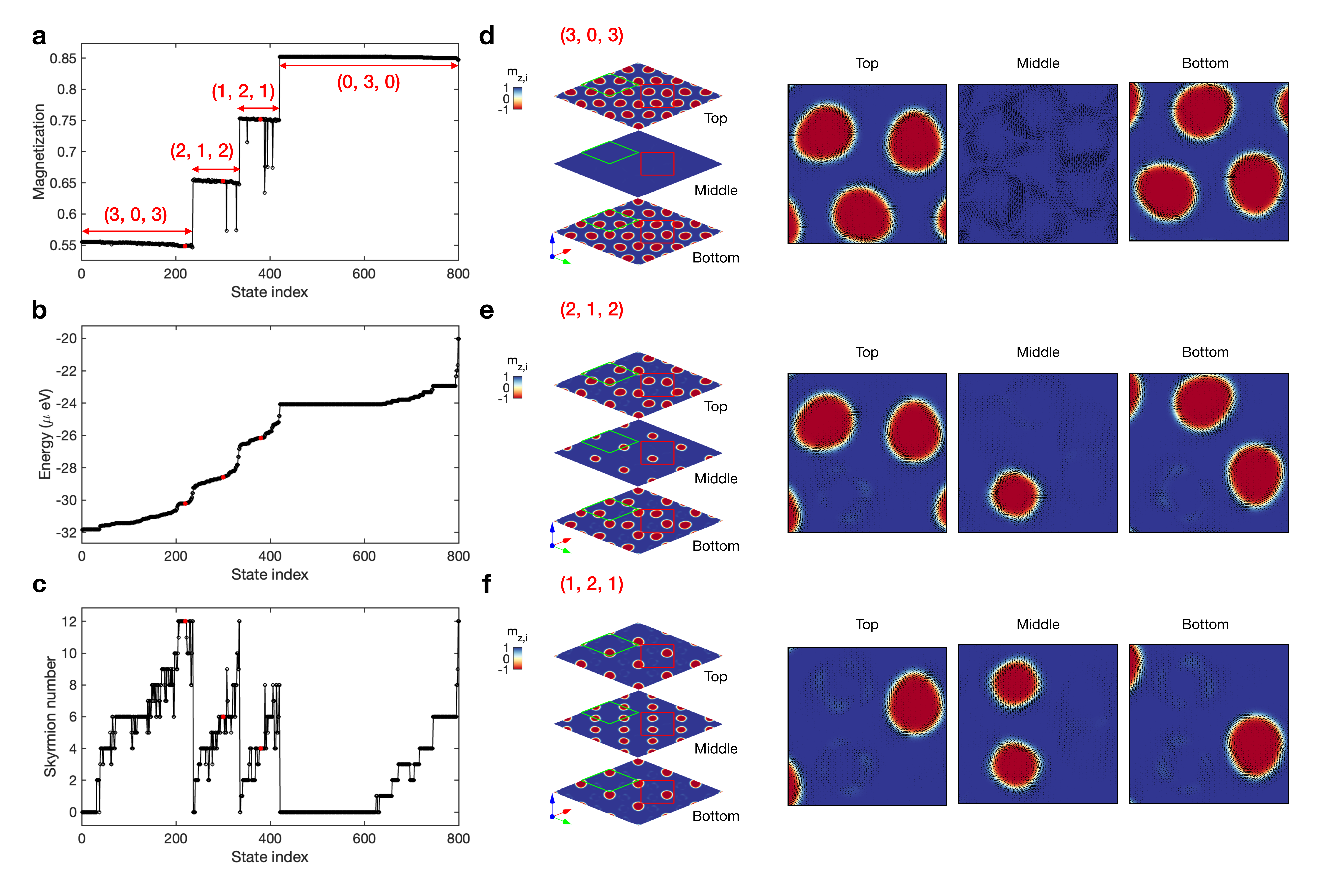}
      \caption{\textbf{Stabilized magnetic states and skyrmion spin textures in the type-II magnetic domain phase.} \textbf{a}-\textbf{c} Out-of-plane magnetization (\textbf{a}), magnetic energy per spin (\textbf{b}), and skyrmion number (\textbf{c}), where the state index denotes stabilized magnetic states from different random initial configurations sorted by energy. In \textbf{a}, the numbers in parentheses indicate the number of magnetic domains in each layer (top, middle, and bottom, respectively). \textbf{c} Magnetic configuration in the excited state with (3,0,3) domains, where the local magnetization in the out-of-plane direction ($m_{z,i}$) is depicted by a color scale. The $3\times3$ moiré superlattice unit cells are shown, where the green line denotes a single unit cell. The second to fourth panels show the magnified area (red line) in each layer in the first panel, where the arrows denote the in-plane components of the magnetization. \textbf{e} Magnetic configuration in the excited state with (2,1,2) domains. \textbf{f} Magnetic configuration in the excited state with (1,2,1) domains. For \textbf{a}-\textbf{f}, we use $J=2$ meV, $D_z=0.2$ meV, and $\theta=1.20^\circ$ [$(m,n)=(27,28)$].}    
	\label{figS3}
\end{figure*}

Figure \ref{figS3} illustrates stabilized magnetic states obtained from various random initial configurations in the type-II magnetic domain (MD-II) phase. In addition to the ground state with (3,0,3) domains, the magnetic domain states carrying different numbers of domains [(2,1,2), (1,2,1), (0,3,0)] are also identified, each having a reduced net magnetization [Fig. \ref{figS3}\textbf{a}]. The creation of domains leads to an increase in magnetic energy, indicating that they are metastable states [Fig. \ref{figS3}\textbf{b}]. Additionally, these magnetic domains contain skyrmion spin textures around their domain wall boundaries [Fig. \ref{figS3}\textbf{c}-\textbf{f}], where an increase in the skyrmion number results in a slight rise in energy.

\end{document}